\documentclass[10pt,a4paper]{article}
\usepackage[utf8x]{inputenc}
\usepackage{ucs,color,enumerate}
\usepackage[english]{babel}
\usepackage{amsmath}
\usepackage{amsfonts}
\usepackage{amssymb}
\usepackage[left=3cm,right=3cm,top=3cm,bottom=3cm]{geometry}
\usepackage[round]{natbib}
\usepackage{graphicx,bbm}
\usepackage{hhline}
\usepackage{subfigure}
\usepackage{bbm}

\newcommand{\waffect}{${\tt waffect}$\ }

\usepackage[colorlinks=true, pdfstartview=FitV, linkcolor=blue, citecolor=red, urlcolor=red]{hyperref}
\definecolor{dred}{rgb}{0.85,0,0} 
\definecolor{dgreen}{rgb}{0,0.7,0} 
\definecolor{dblue}{rgb}{0.1,0.1,0.95}

\title{Non-subjective power analysis to detect G$\times$E interactions in Genome-Wide Association Studies in presence of confounding factor}

\author{F. Alarcon$^1$, V. Perduca$^1$, G. Nuel$^2$}

\begin{document}


\maketitle

\hspace{-0.65cm}
$^1$ MAP5-UMR CNRS 8145, Sorbonne Paris Cité University, Paris, France.\\
$^2$ LPMA, UPMC, Sorbonne University, Paris, France.\\

\begin{abstract}

\textbf{Background.} It is generally acknowledged that most complex diseases are affected in part by interactions between genes and genes and/or between genes and environmental factors. Taking into account environmental exposures and their interactions with genetic factors in genome-wide association studies (GWAS) can help to identify high-risk subgroups in the population and provide a better understanding of the disease. For this reason, many methods have been developed to detect  gene-environment (G$\times$E) interactions. Despite this, few loci that interact with environmental exposures have been identified so far. Indeed, the modest effect of G$\times$E interactions as well as confounding factors entail low statistical power to detect such interactions.

\textbf{Results.}
In this work, we provide a simulated dataset in order to study methods for detecting G$\times$E interactions in GWAS in presence of confounding factor and population structure. Our work applies a recently introduced non-subjective method for H1 simulations called \waffect and exploits the publicly available HapMap project to build a datasets with real genotypes and population structures. We use this dataset to study the impact of confounding factors and compare the relative performance of popular methods such as PLINK, random forests and linear mixed models to detect G$\times$E interactions.

\textbf{Conclusion.}
Presence of confounding factor is an obstacle to detect G$\times$E interactions in GWAS and the approaches considered in our power study all have insufficient power to detect the strong simulated interaction.
Our simulated dataset could help to develop new methods which account for confounding factors through latent exposures in order to improve power.
\end{abstract}

\newpage
\section{Introduction}

Genome-wide association studies (GWAS) are a standard method to identify common genetic factors that influence health and disease conditions. These methods have improved our understanding of the genetic basis of many complex traits and are among the most used tools for analyzing complex diseases.
However, it is known that most complex diseases (e.g. diabetes, asthma and cancer) are due to combined effect of genes, environmental factors, as well as their interactions \citep{murcray2009gene}. 

Over the last years, considerable efforts have been put to detect gene-environment interactions (G$\times$E) in GWAS and few loci that interact with environmental exposures have been identified \citep{rothman2010multi,hamza2011genome, garcia2010analysis}. 

However, this problem is well known to be challenging due, in part, to the modest effect of such interactions in terms of relative risk. 
Another reason why detecting G$\times$E interactions is difficult is population structure which can partially explain spurious associations \citep{astle2009population}.
Environmental confounding factors can also be an obstacle for detecting G$\times$E interactions. VanderWeele et al. have studied extensively the implications of confounding factors in G$\times$E interactions studies  \citep{vanderweele2013environmental, vanderweele2012sensitivity}. However, they were more concerned about the rise of false positives in presence of environmental confounding than by the lose of power to detect the causal loci. Indeed, if the environmental factor that interacts with a susceptibility gene is unobserved but is correlated at some extent with one or several observed factors, not taking it into account could decrease radically the power to detect the G$\times$E interactions.

\vspace{0.3cm}
Nowadays, several methods are available to perform GWAS. In principle, they could be used to detect G$\times$E interactions. Among them, PLINK \citep{purcell2007plink} can be considered as a gold standard for classical analysis. A major concern in GWAS is the need to account for the complicate dependence structure in the data, between loci as well as between individuals. Effects of population stratification can be easily accounted in PLINK by adding the PCA's first components as covariates.

As an alternative, linear mixed models stem as promising statistical methods to correct for the stratification in the population. A popular implementation of linear mixed models is Fast-LMM \citep{lippert2011fast}.

Furthermore, powerful data mining techniques are being increasingly used. Among them, the application of random forests (RFs) to the discovery of SNPs related to human diseases has grown in recent years \citep{goldstein2010application}.

\vspace{0.3cm}
Each time a new method is introduced, it is obviously essential to evaluate its performance in comparison with existing techniques through power studies \citep{spencer2009designing}. However, such studies are often conducted by performing H1 simulations under models which are very similar to the ones used to design the new method, thus giving it an obvious advantage over the other methods. It is hence quite common to see many concurrent approaches each claiming to outperform all others.

Recently, a new method for H1 simulations called \waffect (pronounced "double-u affect" for weighted affectations) has been introduced to avoid this issue \citep{perduca2012alternative}. Indeed, this method does not make any other assumption than the causal disease model itself, whose choice is completely unconstrained. \waffect uses weighted permutations to generate phenotypes conditionally to the genotypes and covariates by taking into account both the causal disease model and the design of the study. With this new approach, it is hence possible to produce non-subjective H1 datasets which do not favor one analysis method over the others.

\vspace{0.3cm}
The aim of this paper is twofold: 1- To propose an useful simulated dataset based on real genotypes with a confounding factor that interacts with the causal locus and is correlated with the observed covariates. For this purpose, we exploited the publicly available HapMap project datasets \citep{thorisson2005international} to obtain real genotypes with population structures, and we used \waffect to generate phenotypes for a chosen causal disease model. This simulated dataset could be very useful for future works aiming at studying the many issues involved in the detection of G$\times$E interactions in GWAS. 
2- To study the effect of a confounding factor on the power to detect G$\times$E interactions in GWAS. For this purpose, we compared four approaches based on three popular methods (PLINK, Fast-LMM and random forests) by performing power analysis on our simulated dataset.

\section{Simulated Dataset}

\subsection{Genotypes}
The genotypic dataset used in our study was extracted from the HapMap phase III database of genetic variations \citep{gibbs2003international}.  This database investigates 11 human populations including 57 unrelated MEX, 146 unrelated YRI, 52 unrelated ASW, 110 unrelated CEU, 154 unrelated MKK, 137 unrelated CHB, 109 unrelated CHD, 101 unrelated GIH, 113 unrelated JPT, 110 unrelated LWK and 102 unrelated TSI. Only the SNPs that are shared by all populations were retained. 

\begin{figure}
\begin{center}
\includegraphics[width=0.8\textwidth]{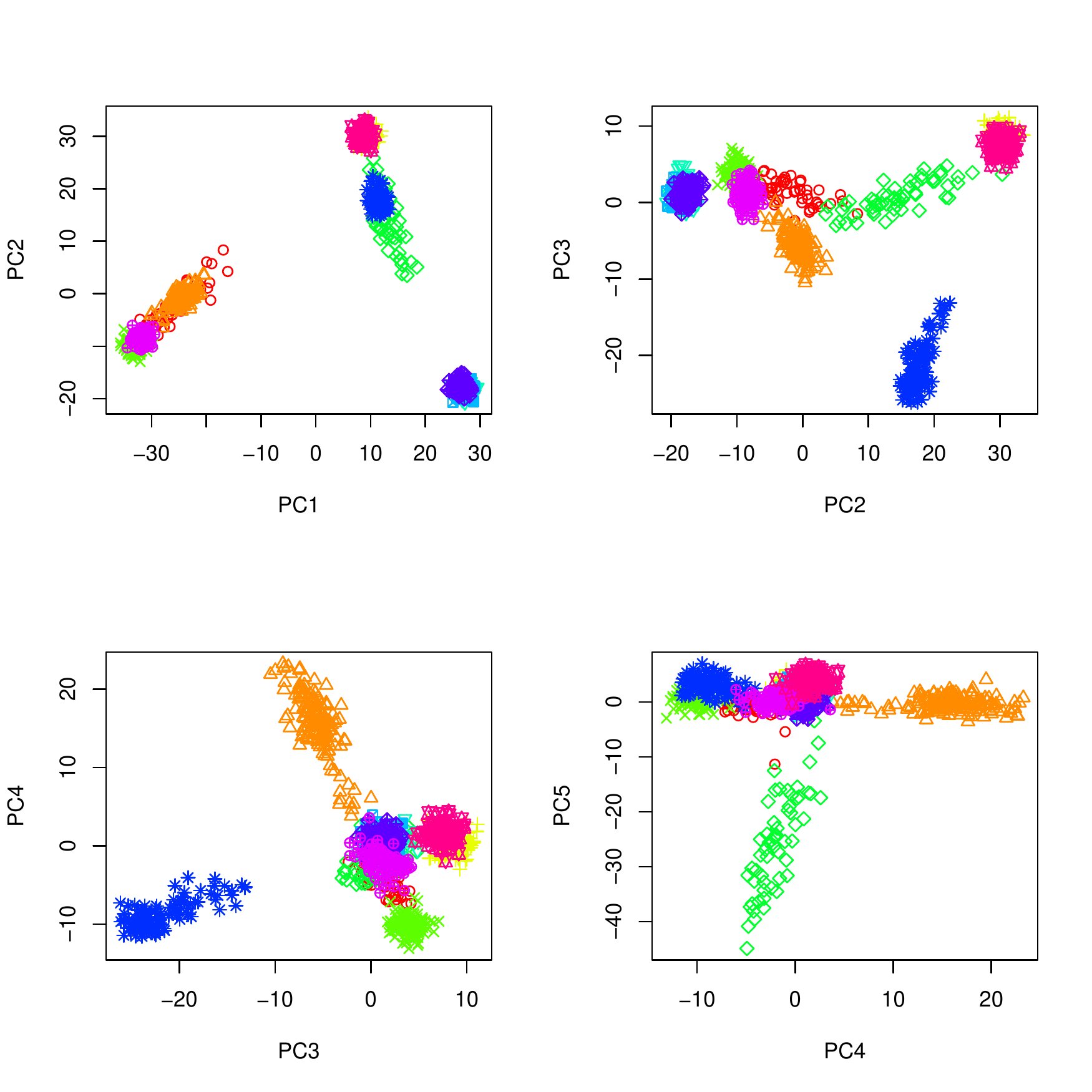} 
\end{center}
\caption{PCA of the genotypes. Scatterplots of the first five principal components. Colors (and symbols) correspond to the 11 human populations in the dataset.}\label{fig-pca}
\end{figure}

Principal component analysis (PCA) was performed on the whole genome, keeping one SNP over 1000 SNPs; the first five principal components (${\tt pca}_{\,i}, i=1,\ldots,5$) were considered as covariates. Figure~\ref{fig-pca} shows that the PCs allow to retrieve easily the information about the population. 

The association analysis in our study were conducted on Chromosome 6, after quality control including Hardy Weinberg equilibrium testing and exclusion of SNPs with a minor allele frequency (MAF) less than 5\%.

\subsection{Covariates}

Covariates and phenotypes were simulated in order to mimic a complex interaction between an arbitrarily chosen causal SNP and a confounding factor through a latent exposure (called ${\tt treatment}$). The latent exposure was defined with high correlation with two observed covariates (${\tt bmi}$, for body mass index, and ${\tt sex}$) as well as with the population of belonging. The idea is that the treatment is typically taken by women (and less often by men) trying to loose weight. 

$\tt{bmi}$ was simulated taken into account the five first principal components and another environmental covariate denoted ${\tt smoking}$. This binary variable was simulated to mimic smoking behaviors with a probability distribution which depends on the population and sex. Indeed, women usually smoke less than men with this difference depending on the population. In order to simulate the smoking covariate, the eleven populations were classified in three sub-populations: European (E); African (Af) and Asian (As). 

In the European sub-population, we considered that 32\% of individuals were smokers. More specifically, we supposed that 37\% of men and 27\% of women were smokers. In the African sub-population, the prevalence of smoking was supposed to be 27\%: 43.8\% among men and 12.9\% among women \citep{christopoulou2011role}. At last, in the Asian sub-population, we considered that 27\% of individuals were smokers: 45.7\% among men and 4.8 \% among women \citep{tsai2008gender}. Covariate ${\tt sex}$ was obtained from HapMap data.

Specifically,  ${\tt bmi}$ was simulated with a regression on the first five principal components in order to have 60\% of heritability and with a residual standard deviation of $4.0$. To take into account the fact that in average smokers have a lower ${\tt bmi}$ than non-smokers, we simulated a smoker effect in the ${\tt bmi}$ covariate by adding a score of 1.5 to the {\tt bmi} average for non-smokers \citep{chiolero2008consequences}.
 
The individual probability to take a treatment, $\mathbb{P}({\tt treatment})$, was correlated with covariates ${\tt bmi}$, ${\tt sex}$ and the population of belonging as follows: 

$$1/\mathbb{P}({\tt treatment}) = (1+2 \times 1_{{\tt sex}=1})\times [1+\exp({-{\tt bmi}+25+ \gamma})]$$
where $\gamma \in \{-\inf,-0.1,0,0.15,-0.45,0.35,0.6,-0.4,0.05,0.1\}$ for population 1 to 11.

\vspace{0.3cm}
To sum up, in our standard design covariates ${\tt sex}$, ${\tt pca}_{\,i}, (i=1,\ldots,5)$, ${\tt bmi}$ and ${\tt smoking}$ were supposed observed. The population of belonging,  covariate ${\tt treatment}$ and the other principal components were supposed unknown (even though these are easily computable).

\subsection{Disease Model}

We arbitrarily chose the SNP in position 22,683,075 in a dense area of chromosome 6, as the binary susceptibility locus, denoted ${\tt causalSNP}$. Assuming a dominant effect, we encoded ${\tt causalSNP} = 1$ in presence of at least one minor frequency allele and ${\tt causalSNP} = 0$ otherwise. We considered a disease model with a very strong G$\times$E interaction (relative risk of $50$) with a baseline prevalence of $1\%$:
$$
\mathbb{P}({\tt disease})=0.01 \times (1.0 + 50.0 \times 1_{{\tt causalSNP}=1} \times 1_{{\tt treatment}=1})
$$

It is important to stress that this is not a very realistic model of complex disease due to the lack of genetic marginal effects and also due to the very strong interaction with a relative risk (RR) equal to 50. We chose not to include any marginal effects for sake of clarity and because we were interested on the detection of G$\times$E interactions. Concerning the strong G$\times$E interaction, we chose to have a relative risk as 50 in order to increase the chance of detecting the G$\times$E interaction in presence of the simulated confounding factor. 

Phenotypes were simulated accordingly to the disease models by means of the package \waffect \citep{perduca2012alternative} publicly available on the CRAN server of R packages \citep{R2013}. This enabled us to simulate exactly 595 cases and 596 controls for the 1191 individuals from the HapMap genotypic dataset (see next session for a comprehensive introduction to \waffect).

%

\section{Power analysis}

\subsection{Phenotype Simulations}

In order to assess the empirical statistical power of different tools to detect associations, we simulated 200 phenotypes replicates under the disease model H1 and 200 phenotypes replicates under the null hypothesis H0 of no association. Each replicate consists of 1191 phenotypes, one for each individual.

The simulations under H0 were obtained by simply permuting the phenotypes, thus breaking potential associations between phenotypes and genotypes. The simulations under the alternative hypothesis H1 were performed using the R package \waffect publicly available on CRAN \citep{perduca2012alternative}. 

The principal function in \waffect is based on a backward sampling algorithm which makes it possible to generate weighted permutations. For the purposes of phenotype simulation, the vector of weights is given by the penetrance, that is the probability for each individual to be a case according to the disease model. One crucial consequence of using weighted permutations is that the number of cases and controls is constant across the replicates. This makes it possible to respect the original design in each replicate and therefore to compare the performance of an association method across different replicates. 

Simulating phenotypes rather than genotypes, as does the gold standard Hapgen \citep{su2011hapgen2}, does not require additional data such as haplotype frequencies or recombination rates and has the obvious advantages of requiring much smaller computational memory and time. Moreover, for the purposes of the present work, the primary benefit of using \waffect is that it only requires a vector of probabilities as input. As a result, the choice of the disease model is totally unconstrained; in particular it is possible to include G$\times$E interactions. 

In principle, one could achieve the same result by simply using a rejection algorithm which samples the phenotype of each individual according to the probability to be a case and then accepts the resulting replicate only if there are \emph{enough} cases. Because the probability of obtaining a full configuration of phenotypes with the correct number of cases is  extremely low, this approach cannot be used in practice. In order to overcome this problem, a solution often applied in practice is to increase the prevalence in the disease model, maintaining unchanged the relative risks. However it can be proved that adjusting the prevalence creates bias in the empirical power estimate \citep{perduca2012alternative}.

\subsection{Statistical analysis}

In this section, we briefly describe the four popular approaches adopted in our study to perform GWA analysis. The goal standard PLINK \citep[individual SNP logistic regression,][]{purcell2007plink} was applied in two alternative approaches :  1- analysis performed regardless of G$\times$E interactions and considering only genetic effects; and, since it is easy to consider interaction terms with PLINK, 2- analysis performed taking into account G$\times$E interactions. The other two approaches are 3- the linear mixed model for population structure correction implemented in Fast-LMM \citep{lippert2011fast}, and 4- the random forest  data mining technique \citep[RandomForest R package,][]{liaw2002rf}.

\subsubsection{PLINK}\label{sec:plink}

PLINK implements a logistic regression approach \citep{purcell2007plink} that allows for multiple binary or continuous covariates when testing for disease trait SNP association and interactions with covariates.
PLINK provides p-values for significance coefficients in the logistic model. In this work, we considered two approaches using PLINK. 

The first approach, which we referred to as "PLINK SNP", consisted in performing analysis regardless of G$\times$E interactions by looking at the p-value associated to the significant coefficients for the SNPs. 

We referred to the second approach as "PLINK SNP$\times$COV", where COV is the environmental covariate under consideration (either $\tt{bmi}$ in our standard design or $\tt{treatment}$ in our further analysis, see below). PLINK SNP$\times$COV accounted for all the interactions between the analyzed SNPs and the environmental factor considered through the p-values associated to the significance coefficients of such interactions.

Correction for population structure was taken into account by considering the five first principal components resulting from the PCA performed on the whole genome.

\subsubsection{Random Forests}\label{sec:rf}

Random forests (RFs) have been introduced by \citet{breiman2001random}. The general principle consists in building repeatedly classification and regression trees (CART) from bootstrapping of the original data. This process produces a forest of classification trees which are statistically analyzed to produce importance measures of the covariates (e.g. a variable belonging to many trees probably plays a key role in the classification). 

Random forests are a popular way to perform data mining on GWAS data. Despite the fact that they exploit heavily marginal linear regressions, random forests are able to detect interactions between variables \citep[see][for an overview of random forests in the GWAS context]{boulesteix2012overview}. Recently, a regularized version of random forests has been proposed to deal with high dimensional data \citep{deng2012rrf}. In this work we decided to disregard this approach because it was too slow on our data to be practical.

For our random forests analysis, we used the {\tt randomForest} package (version 4.6-7) from R \citep{R2013}. We simply used the default parameters of the method with the disease status as a binary outcome and with all observed covariates and SNPs as explanatory variables. Once the forest computed, we extracted for each variable its importance measure using the default approach of the package (normalized difference between out-of-bag proportion error using original data or a permuted version). We hence obtained for each variable and each replicate a real value which reflects the importance of the variable for discriminating between cases and controls. The higher this importance value, the stronger the association with the disease.

\subsubsection{Fast-LMM}\label{sec:lmm}

It is a well known problem that in GWAS confounding effects of population structure lead to false positive and therefore need to be taken into account. An alternative to including the first principal components in linear or logistic regression models in order to correct for confounding factors are Linear Mixed Models \citep[LMMs,][]{10.1371/journal.pone.0075707}. LMMs generalize linear models by introducing random effects as predictors, in addition to the usual fixed effects. Indeed, LMMs are known to be effective when observations are not independent but rather involve related individuals.

In LMMs for GWAS, the random effect is expressed by a multivariate normal distribution whose variance-covariance matrix measures the genetic similarity between individuals. Recently the algorithm Fast-LMM has been introduced to perform efficiently exact inference for LMMs \citep{lippert2011fast}. Roughly speaking, Fast-LMM (for  Factored spectrally transformed Linear Mixed Models) is based on a spectral decomposition of the genetic similarity matrix which rotates the phenotypes into uncorrelated phenotypes thus converting the original estimation problem into the maximization of the likelihood of a linear regression model.  

A drawback of the current implementation of Fast-LMM is that it does not allow to consider explicit interaction terms between genotypic variables and covariates in the linear mixed model \footnote{(We unsuccessfully tried to contact the authors of Fast-LMM on this matter.)}. A possible solution to overcome this limitation is to code directly such interactions in the covariate file, thus adding new columns. However, this solution was not appropriate in this context because it would have required to magnify several times the size of the variables file in order to consider the cartesian product of all the SNPs with all the covariates. We then we simply applied Fast-LMM to the original genotypic and covariate datasets.

\subsection{Power, ROC curves and AUC}

In order to evaluate empirically the performance of the methods described above, we computed four summary statistics, one for each method. These global statistics were then used to estimate the Area under the Curve (AUC) corresponding to the receiver operating characteristic (ROC) curves of the four methods, each expressing the performance of the corresponding method. 

More specifically, for PLINK SNP we took as simple global statistics the smallest among all the p-values associated to the significance coefficients of the SNPs, similarly for Fast-LMM. For PLINK SNP$\times$COV, we took the smallest among all the p-values associated to significance coefficients of the terms coding for the interactions between the SNPs and the covariate ({\tt bmi} or {\tt treatment}). At last, the summary statistics for the random forest-based method was defined as the maximum of the importance statistics over all considered SNPs.

Then, for each method, we obtained two vectors of length 200,  one under H0 and one under H1. These two vectors of comprehensive signals were used to estimate the ROC curves of the four methods using the R package pROC \citep{robin2011proc}.

We recall that ROC curves provide a graphical representation of the specificities and sensitivities (i.e. values of statistical power) that can be obtained for all possible values of the threshold of significance \citep{metz1978basic}. An informative summary of the ROC curve information is the area under the ROC curve (i.e. AUC). The AUC can be qualitatively interpreted as follows: $\text{AUC} \leq 0.6$ means ‘fail’; $0.6 < \text{AUC} \leq 0.70$ means ‘poor’; $0.7 < \text{AUC} \leq 0.80$ means ‘fair’; $0.8 < \text{AUC} \leq 0.9$ means ‘good’; $0.9 < \text{AUC} \leq 1.0$ means ‘excellent’.

\section{Results and discussion}

Association analysis were adjusted on covariates $\tt{sex}$, $\tt{smoking}$ together with either {\tt bmi} (our standard design) or {\tt treatment}. Moreover, for PLINK and random forests the five first principal components $\tt{pc_i, i\in \{1, \dots, 5\}}$ were included as predictors. We recall that in order to evaluate empirically the detection power of the four approaches in presence of confounding factor, analysis were performed on 200+200 phenotypic replicates under H0 and H1. Investigations were performed on chromosome 6 and subregions. 

\subsection{Confounding factor is not known}
At first, the latent exposure $\tt{treatment}$ (i.e. the confounding factor) was supposed to be unknown :  the covariate $\tt{bmi}$ was observed instead of  $\tt{treatment}$ and considered, mistakenly, as the environmental exposure of interest. 

Table~\ref{tab-auc-bmi} shows the estimated AUC together with 95\% confidence intervals for the four approaches. Obviously, the performance of each method increases when the region under consideration decreases and reach a good power when the region is restricted to the causal SNP.  However power is low (fail or poor) when estimation are done on whole chromosome 6.
PLINK SNP, Fast-LMM and RF provide comparable results even if PLINK SNP seems to be a little more efficient. 

Surprisingly, the estimated performance is better with PLINK SNP than with PLINK SNP$\times \tt{bmi}$. For example, applying PLINK SNP on a region with less than 800 SNPs around the causal SNP provides good to excellent power while analysis with PLINK SNP$\times \tt{bmi}$ need to be restricted to the causal SNP to reach similar power. At first, this counter-intuitive result led us to believe that it is unnecessary to account for the G$\times$E interactions in analysis since PLINK SNP provides better power than PLINK SNP$\times \tt{bmi}$.

Furthermore, we note that AUC estimate applying PLINK SNP$\times \tt{bmi}$ when the analysis is performed on the whole chromosome 6 appears higher than on a region of 4000 SNPs around the causal SNP. This result could be explained by sample variability.


\subsection{Confounding factor is known}

To try to understand the fact that taking into account G$\times$E interactions in presence of a confounding factor seems not to improve the performance, we performed association analysis by replacing the covariate $\tt{bmi}$ by the confounding factor $\tt{treatment}$. This is possible because analysis are done on our simulated dataset for which we know the covariate $\tt{treatment}$. 

Table~\ref{tab-auc-treatment} shows the results obtained observing the confounding factor. In this context, AUC estimated with PLINK SNP$\times \tt{treatment}$ are equal to 1, as expected given the strength of the simulated G$\times$E interaction. This results show that the approach accounting for the interaction is better than approach accounting only for the SNP and demonstrate  the importance of accounting for G$\times$E interaction in a process of consideration of a confounding factor.

In contrast, the AUC estimated with PLINK SNP are almost equal to the AUC estimated when the confounding factor $\tt{treatment}$ is latent and instead the covariate $\tt{bmi}$ is observed. As expected, Fast-LMM gives similar results to PLINK SNP. 
For the RF, we can only observe a slight improvement ($1$ to $3\%$ of AUC). This is due to the fact that, like for Fast-LMM or PLINK SNP, the RF approach does not consider explicitly interactions with the covariates. However, RF are known to be able to capture complex non-linear interaction between covariates which tend to be co-selected in the same trees. In our example, this alleged feature clearly shows its limits.

\subsection{Further considerations}

An interesting property of Fast-LMM is it accounts for population structure by introducing random effects as predictors. In this context, analysis were performed with Fast-LMM, with no principal components as covariates. We verified (results not shown) that accounting  explicitly for principal components gives similar results. In the same way, we performed analysis with PLINK SNP$\times \tt{bmi}$ considering explicitly the population of belonging instead of principal components and found similar results than considering principal components (results not shown). 

\begin{figure}[!htp]
\begin{center}
\subfigure[Observing the covariate $\tt{treatment}$.]{\label{plink-manhattan-plots-treat} \includegraphics[width=0.45\textwidth]{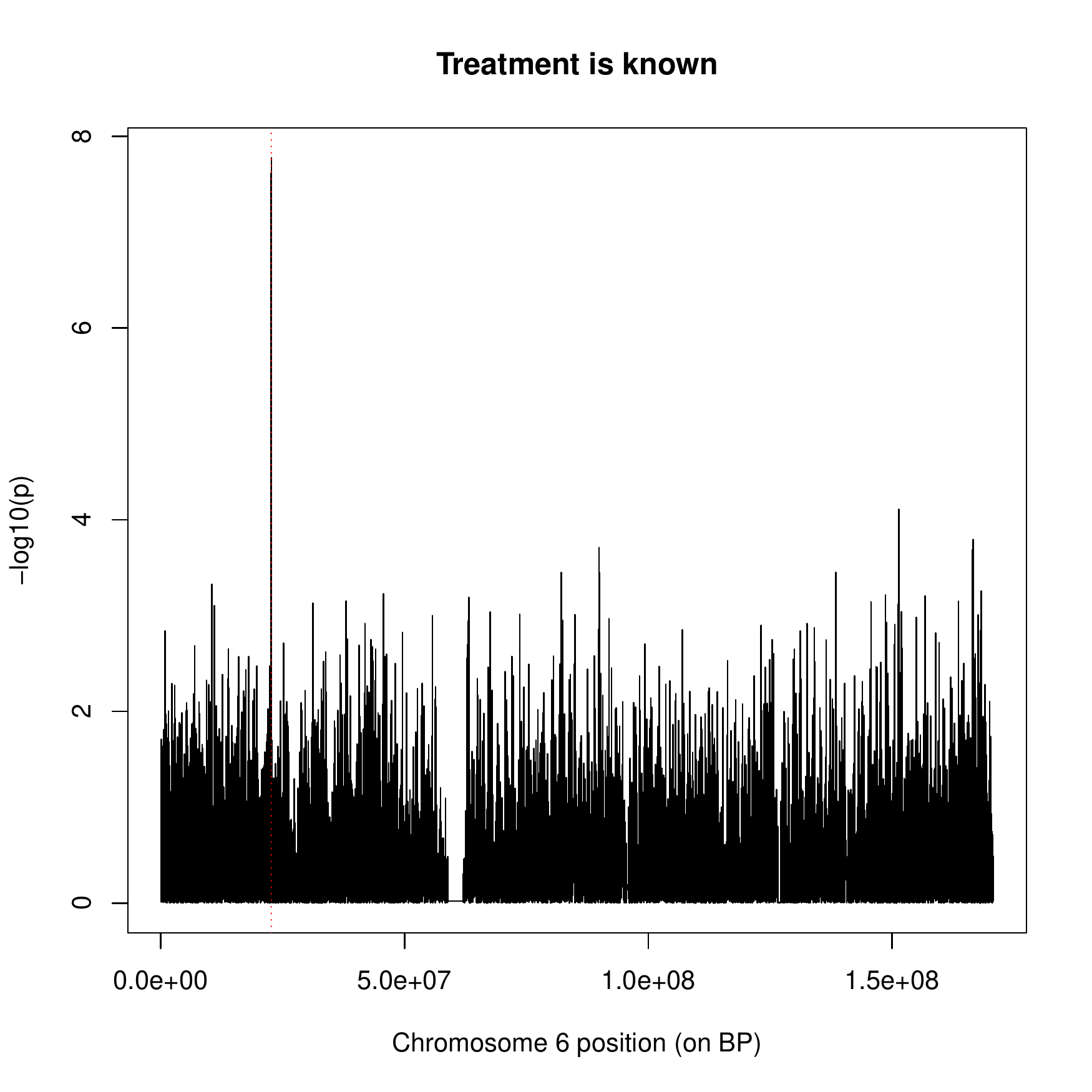}}
\subfigure[Observing the covariate $\tt{bmi}$. ]{\label{plink-manhattan-plots-bmi} \includegraphics[width=0.45\textwidth]{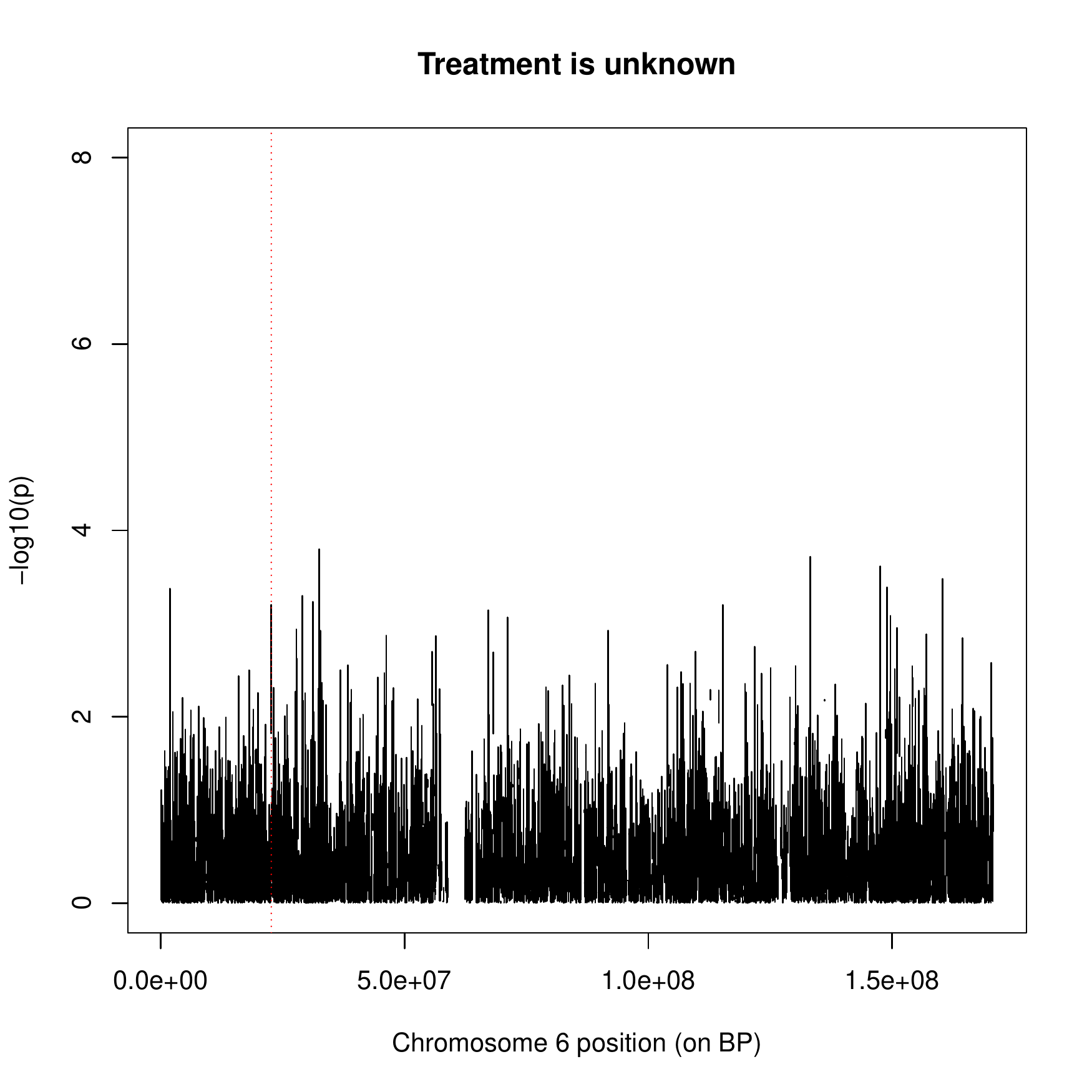}}
\caption{Manhattan plots considering Chromosome 6 and observing the covariate $\tt{treatment}$ or the covariate $\tt{bmi}$. The red vertical line indicates the causal SNP}
\label{plink-manhattan-plots}
\end{center}
\end{figure}

Another question that may arise is about the signal localization.  Indeed, presence of confounding factor makes very difficult to identify the causal SNP. Figure~\ref{plink-manhattan-plots} shows the Manhattan plots obtained performing PLINK SNP$\times$COV analysis on Chromosome 6 from one simulation under H1. The vertical red line indicates the causal SNP location. On figure~\ref{plink-manhattan-plots-treat} analysis was performed observing the confounding factor $\tt{treatment}$ (i.e. using PLINK SNP$\times \tt{treatment}$ approach) and on figure ~\ref{plink-manhattan-plots-bmi} analysis was performed in presence of confounding factor. (i.e. using PLINK SNP$\times \tt{bmi}$ approach). When the confounding factor is known, the location detected by the signal is the same as the causal SNP location. In contrast, in presence of confounding factor,  there is no signal clearly detected.

\newpage			

\section{Conclusion}

This article presents a simulated dataset - based on the HapMap project data with real genotypes and population structures - which can be helpful in studying the impact of a confounding factor on the detection of G$\times$E interactions in GWAS. Furthermore, the proposed dataset was used to conduct a non-subjective power analysis using three popular GWAS statistical methods (PLINK, Random Forests and Fast-LMM) and four corresponding summary statistics. The non-subjectivity comes from the fact that the phenotypes under H1 as well as under H0 were simulated with \waffect which does not require any additional information other than the original genotypes. The dataset will be made available on ENA at the time of publication, so that all replication will be feasible.

The disease model was built with a strong interaction between the causal SNP (which was chosen arbitrarily) and an unobserved environmental exposure to a treatment and no main effect given by the causal SNP or the latent exposure. Obviously, the high relative risk resulting from the considered G$\times$E interaction and the absence of a marginal SNP effect are unrealistic. However, the confounding factor causes such a loss of detection power that it is necessary to study such typical case in first place. Indeed, analysis performed on the whole chromosome 6 shows a poor signal detection power for the four approaches considered in presence of confounding factor. Furthermore, we show that when analysis is performed regardless the interaction, the fact to observe or not the confounding factor $\tt{treatment}$ has no impact on the detection power. These results highlight the importance of taking into account the G$\times$E interactions at the risk of finding no signal at all.

In this work, we chose to focus on three popular methods belonging to different families of statistical techniques. 
Unfortunately, the current implementation of Fast-LMM does not allow to account for G$\times$E interactions. The proposed dataset has the potential to provide a good framework to develop further features of Fast-LMM enabling it to account for such interactions. As an alternative, we considered the idea to precompute a full covariate matrix including interaction to Fast-LMM, but this approach was finally discarded since there was no way to force fast-LMM to perform the appropriate significance tests. Similar consideration hold for the RF which are not specifically designed to deal with explicit interactions.

Other methods based on the linear model are known for case control data, in order to detect G$\times$E interactions. Kraft and al. \citep{kraft2007exploiting} proposed a powerful 2-df joint test of marginal association and G$\times$E interaction. Shortly after, Mucray and al. \citep{murcray2009gene} proposed a 2-step approach for detecting G$\times$E interaction in GWA studies. 
Dai and al.\citep{dai2012simultaneously} proposed a new way to combine the test of marginal genetic effect and the test of G$\times$E interaction, by exploiting the independence between the two tests.
While these methods have demonstrated their efficiency, their performance was assessed through simulations that do not  account for realistic complexity such as the inclusion of confounding factor and/or of complicated dependance structures between individuals as well as between loci. 
Testing the detection power of these tests on our realistic simulated dataset could then be an interesting development. Results obtained in this paper already seem to show that methods based on the linear model have poor power. As a consequence, we can expect that the tests mentioned above will not perform well either. 

In conclusion, efforts should be put in developing methods able to take into account confounding factors such as a potential latent exposure.

\newpage
\bibliographystyle{plainnat}
\bibliography{gwas_abstract_biblio}

\begin{thebibliography}{28}
\providecommand{\natexlab}[1]{#1}
\providecommand{\url}[1]{\texttt{#1}}
\expandafter\ifx\csname urlstyle\endcsname\relax
  \providecommand{\doi}[1]{doi: #1}\else
  \providecommand{\doi}{doi: \begingroup \urlstyle{rm}\Url}\fi

\bibitem[Astle and Balding(2009)]{astle2009population}
William Astle and David~J Balding.
\newblock Population structure and cryptic relatedness in genetic association
  studies.
\newblock \emph{Statistical Science}, 24\penalty0 (4):\penalty0 451--471, 2009.

\bibitem[Boulesteix et~al.(2012)Boulesteix, Janitza, Kruppa, and
  K{\"o}nig]{boulesteix2012overview}
Anne-Laure Boulesteix, Silke Janitza, Jochen Kruppa, and Inke~R K{\"o}nig.
\newblock Overview of random forest methodology and practical guidance with
  emphasis on computational biology and bioinformatics.
\newblock \emph{Wiley Interdisciplinary Reviews: Data Mining and Knowledge
  Discovery}, 2\penalty0 (6):\penalty0 493--507, 2012.

\bibitem[Breiman(2001)]{breiman2001random}
Leo Breiman.
\newblock Random forests.
\newblock \emph{Machine learning}, 45\penalty0 (1):\penalty0 5--32, 2001.

\bibitem[Chiolero et~al.(2008)Chiolero, Faeh, Paccaud, and
  Cornuz]{chiolero2008consequences}
Arnaud Chiolero, David Faeh, Fred Paccaud, and Jacques Cornuz.
\newblock Consequences of smoking for body weight, body fat distribution, and
  insulin resistance.
\newblock \emph{The American journal of clinical nutrition}, 87\penalty0
  (4):\penalty0 801--809, 2008.

\bibitem[Christopoulou and Lillard(2011)]{christopoulou2011role}
Rebekka Christopoulou and Dean~R Lillard.
\newblock The role of culture in smoking behavior: evidence from british
  immigrants in australia, south africa, and the us.
\newblock Technical report, Cornell University, 2011.

\bibitem[Dai et~al.(2012)Dai, Logsdon, Huang, Hsu, Reiner, Prentice, and
  Kooperberg]{dai2012simultaneously}
James~Y Dai, Benjamin~A Logsdon, Ying Huang, Li~Hsu, Alexander~P Reiner, Ross~L
  Prentice, and Charles Kooperberg.
\newblock Simultaneously testing for marginal genetic association and
  gene-environment interaction.
\newblock \emph{American journal of epidemiology}, 176\penalty0 (2):\penalty0
  164--173, 2012.

\bibitem[Deng and Runger(2012)]{deng2012rrf}
Houtao Deng and George Runger.
\newblock Feature selection via regularized trees.
\newblock \emph{The 2012 International Joint Conference on Neural Networks
  (IJCNN)}, 2012.

\bibitem[Garcia-Closas et~al.(2010)Garcia-Closas, Jacobs, Kraft, and
  Chatterjee]{garcia2010analysis}
M~Garcia-Closas, K~Jacobs, P~Kraft, and N~Chatterjee.
\newblock Analysis of epidemiologic studies of genetic effects and
  gene-environment interactions.
\newblock \emph{IARC scientific publications}, 163\penalty0 (163):\penalty0
  281--301, 2010.

\bibitem[Gibbs et~al.(2003)Gibbs, Belmont, Hardenbol, Willis, Yu, Yang, Ch'ang,
  Huang, Liu, Shen, et~al.]{gibbs2003international}
Richard~A Gibbs, John~W Belmont, Paul Hardenbol, Thomas~D Willis, Fuli Yu,
  Huanming Yang, Lan-Yang Ch'ang, Wei Huang, Bin Liu, Yan Shen, et~al.
\newblock The international hapmap project.
\newblock \emph{Nature}, 426\penalty0 (6968):\penalty0 789--796, 2003.

\bibitem[Goldstein et~al.(2010)Goldstein, Hubbard, Cutler, and
  Barcellos]{goldstein2010application}
Benjamin~A Goldstein, Alan~E Hubbard, Adele Cutler, and Lisa~F Barcellos.
\newblock {A}n application of {R}andom {F}orests to a genome-wide association
  dataset: {M}ethodological considerations \& new findings.
\newblock \emph{BMC genetics}, 11\penalty0 (1):\penalty0 49, 2010.

\bibitem[Hamza et~al.(2011)Hamza, Chen, Hill-Burns, Rhodes, Montimurro, Kay,
  Tenesa, Kusel, Sheehan, Eaaswarkhanth, et~al.]{hamza2011genome}
Taye~H Hamza, Honglei Chen, Erin~M Hill-Burns, Shannon~L Rhodes, Jennifer
  Montimurro, Denise~M Kay, Albert Tenesa, Victoria~I Kusel, Patricia Sheehan,
  Muthukrishnan Eaaswarkhanth, et~al.
\newblock Genome-wide gene-environment study identifies glutamate receptor gene
  grin2a as a parkinson's disease modifier gene via interaction with coffee.
\newblock \emph{PLoS genetics}, 7\penalty0 (8):\penalty0 e1002237, 2011.

\bibitem[Hoffman(2013)]{10.1371/journal.pone.0075707}
Gabriel~E. Hoffman.
\newblock Correcting for population structure and kinship using the linear
  mixed model: Theory and extensions.
\newblock \emph{PLoS ONE}, 8\penalty0 (10):\penalty0 e75707, 10 2013.
\newblock \doi{10.1371/journal.pone.0075707}.
\newblock URL \url{http://dx.doi.org/10.1371%2Fjournal.pone.0075707}.

\bibitem[Kraft et~al.(2007)Kraft, Yen, Stram, Morrison, and
  Gauderman]{kraft2007exploiting}
Peter Kraft, Y-C Yen, Daniel~O Stram, John Morrison, and W~James Gauderman.
\newblock Exploiting gene-environment interaction to detect genetic
  associations.
\newblock \emph{Human heredity}, 63\penalty0 (2):\penalty0 111--119, 2007.

\bibitem[Liaw and Wiener(2002)]{liaw2002rf}
Andy Liaw and Matthew Wiener.
\newblock Classification and regression by randomforest.
\newblock \emph{R News}, 2\penalty0 (3):\penalty0 18--22, 2002.
\newblock URL \url{http://CRAN.R-project.org/doc/Rnews/}.

\bibitem[Lippert et~al.(2011)Lippert, Listgarten, Liu, Kadie, Davidson, and
  Heckerman]{lippert2011fast}
Christoph Lippert, Jennifer Listgarten, Ying Liu, Carl~M Kadie, Robert~I
  Davidson, and David Heckerman.
\newblock Fast linear mixed models for genome-wide association studies.
\newblock \emph{Nature Methods}, 8\penalty0 (10):\penalty0 833--835, 2011.

\bibitem[Metz(1978)]{metz1978basic}
Charles~E Metz.
\newblock Basic principles of roc analysis.
\newblock In \emph{Seminars in nuclear medicine}, volume~8, pages 283--298.
  Elsevier, 1978.

\bibitem[Murcray et~al.(2009)Murcray, Lewinger, and Gauderman]{murcray2009gene}
Cassandra~E Murcray, Juan~Pablo Lewinger, and W~James Gauderman.
\newblock {G}ene-environment interaction in genome-wide association studies.
\newblock \emph{American journal of epidemiology}, 169\penalty0 (2):\penalty0
  219--226, 2009.

\bibitem[Perduca et~al.(2012)Perduca, Sinoquet, Mourad, and
  Nuel]{perduca2012alternative}
Vittorio Perduca, Christine Sinoquet, Rapha{\"e}l Mourad, and Gregory Nuel.
\newblock {A}lternative {M}ethods for {H1} {S}imulations in {G}enome-{W}ide
  {A}ssociation {S}tudies.
\newblock \emph{Human Heredity}, 73\penalty0 (2):\penalty0 95--104, 2012.

\bibitem[Purcell et~al.(2007)Purcell, Neale, Todd-Brown, Thomas, Ferreira,
  Bender, Maller, Sklar, De~Bakker, Daly, et~al.]{purcell2007plink}
Shaun Purcell, Benjamin Neale, Kathe Todd-Brown, Lori Thomas, Manuel~AR
  Ferreira, David Bender, Julian Maller, Pamela Sklar, Paul~IW De~Bakker,
  Mark~J Daly, et~al.
\newblock {PLINK}: a tool set for whole-genome association and population-based
  linkage analyses.
\newblock \emph{The American Journal of Human Genetics}, 81\penalty0
  (3):\penalty0 559--575, 2007.

\bibitem[{R Core Team}(2013)]{R2013}
{R Core Team}.
\newblock \emph{R: A Language and Environment for Statistical Computing}.
\newblock R Foundation for Statistical Computing, Vienna, Austria, 2013.
\newblock URL \url{http://www.R-project.org}.

\bibitem[Robin et~al.(2011)Robin, Turck, Hainard, Tiberti, Lisacek, Sanchez,
  and M{\"u}ller]{robin2011proc}
Xavier Robin, Natacha Turck, Alexandre Hainard, Natalia Tiberti,
  Fr{\'e}d{\'e}rique Lisacek, Jean-Charles Sanchez, and Markus M{\"u}ller.
\newblock proc: an open-source package for r and s+ to analyze and compare roc
  curves.
\newblock \emph{BMC bioinformatics}, 12\penalty0 (1):\penalty0 77, 2011.

\bibitem[Rothman et~al.(2010)Rothman, Garcia-Closas, Chatterjee, Malats, Wu,
  Figueroa, Real, Van Den~Berg, Matullo, Baris, et~al.]{rothman2010multi}
Nathaniel Rothman, Montserrat Garcia-Closas, Nilanjan Chatterjee, Nuria Malats,
  Xifeng Wu, Jonine~D Figueroa, Francisco~X Real, David Van Den~Berg, Giuseppe
  Matullo, Dalsu Baris, et~al.
\newblock A multi-stage genome-wide association study of bladder cancer
  identifies multiple susceptibility loci.
\newblock \emph{Nature genetics}, 42\penalty0 (11):\penalty0 978--984, 2010.

\bibitem[Spencer et~al.(2009)Spencer, Su, Donnelly, and
  Marchini]{spencer2009designing}
Chris~CA Spencer, Zhan Su, Peter Donnelly, and Jonathan Marchini.
\newblock Designing genome-wide association studies: sample size, power,
  imputation, and the choice of genotyping chip.
\newblock \emph{PLoS genetics}, 5\penalty0 (5):\penalty0 e1000477, 2009.

\bibitem[Su et~al.(2011)Su, Marchini, and Donnelly]{su2011hapgen2}
Zhan Su, Jonathan Marchini, and Peter Donnelly.
\newblock {HAPGEN2: simulation of multiple disease SNPs}.
\newblock \emph{Bioinformatics}, 27\penalty0 (16):\penalty0 2304--2305, 2011.

\bibitem[Thorisson et~al.(2005)Thorisson, Smith, Krishnan, and
  Stein]{thorisson2005international}
Gudmundur~A Thorisson, Albert~V Smith, Lalitha Krishnan, and Lincoln~D Stein.
\newblock The international hapmap project web site.
\newblock \emph{Genome research}, 15\penalty0 (11):\penalty0 1592--1593, 2005.

\bibitem[Tsai et~al.(2008)Tsai, Tsai, Yang, and Kuo]{tsai2008gender}
Yi-Wen Tsai, Tzu-I Tsai, Chung-Lin Yang, and Ken~N Kuo.
\newblock Gender differences in smoking behaviors in an asian population.
\newblock \emph{Journal of Women's Health}, 17\penalty0 (6):\penalty0 971--978,
  2008.

\bibitem[VanderWeele et~al.(2012)VanderWeele, Mukherjee, and
  Chen]{vanderweele2012sensitivity}
Tyler~J VanderWeele, Bhramar Mukherjee, and Jinbo Chen.
\newblock Sensitivity analysis for interactions under unmeasured confounding.
\newblock \emph{Statistics in medicine}, 31\penalty0 (22):\penalty0 2552--2564,
  2012.

\bibitem[VanderWeele et~al.(2013)VanderWeele, Ko, and
  Mukherjee]{vanderweele2013environmental}
Tyler~J VanderWeele, Yi-An Ko, and Bhramar Mukherjee.
\newblock Environmental confounding in gene-environment interaction studies.
\newblock \emph{American Journal of Epidemiology}, 2013.

\end{thebibliography}

\newpage

\begin{table}
\begin{center}
\small
\begin{tabular}{ccccc}
\hline
AUC (\%) & PLINK SNP & PLINK SNP$\times$bmi & RF & Fast-LMM \\
\hline
whole chromosome 6 &$64.69\ [59.26-70.13]$ & $56.39\ [50.69-62.08]$ & $66.23\ [60.93-71.55]$ &$61.91\ [56.42-67.41]$  \\
8 000 SNPs region &$72.04\ [67.03-77.05]$ & $55.32\ [49.60-61.04]$ & $71.99\ [67.07-76.91]$  & $68.96\ [63.84-74.07]$\\
2 000 SNPs region &$74.44\ [69.52-79.35]$ &  $58.05\ [52.41-63.70]$ &  $76.15\ [71.49-80.82]$ &$71.36\ [66.35-76.36]$\\
800 SNPs regions &$81.78\ [77.54-86.02]$&  $60.24\ [54.66-65.81]$ & $79.48\ [75.10-83.87]$ &$80.65\ [76.48-84.82] $\\
200 SNPs region &$88.50\ [85.27-91.72]$ & $68.62\ [63.38-73.85]$ &  $84.71\ [80.77-88.66]$ &$86.72\ [83.28-90.17]$ \\
causal SNP & $99.15\ [98.57-99.73]$& $88.67\ [85.48-91.86]$ & $89.03\ [85.75-92.32]$ &$99.09\ [98.49-99.70] $\\
\hline
\end{tabular}
\end{center}
\caption{\label{tab-auc-bmi} Association analysis performed on Chromosome 6 with the four approaches, in presence of a confounding factor (i.e. covariate $\tt{bmi}$ is observed but covariate $\tt{treatment}$ is unknown) . Restricted regions are centered on causal SNP}
\end{table}

\newpage
\begin{table}
\begin{center}
\small
\begin{tabular}{ccccc}
\hline
AUC (\%) & PLINK SNP & PLINK SNP$\times$bmi & RF & Fast-LMM \\
\hline
whole chromosome 6 &$63.35\ [57.84 -68.85 ] $ & $99.97\ [99.93 - 100.0] $ & $67.66\ [62.44 - 72.88 ]$ & $61.54\ [56.02-67.06]$  \\
8 000 SNPs region &$70.67\ [65.58 -75.75 ] $ & $100.0\ [99.99-100.0] $ & $78.30\ [73.88- 82.73]$  &$68.02\ [62.82-73.23]$ \\
2 000 SNPs region & $73.91\ [68.95 -78.88 ] $&$100.0\ [99.99-100.0] $ &  $82.57\ [78.60 - 86.54 ]$ &$72.03\ [67.02-77.04]$ \\
800 SNPs regions &$80.24\ [75.78 -84.69 ] $ & $100.0\ [100.0-100.0]$  & $86.07\ [82.54-89.61]$ &$80.22\ [75.95-84.48]$ \\
200 SNPs region &$87.62\ [84.26 -90.98 ] $ & $100.0\ [100.0-100.0]$ &  $89.88\ [86.87-92.90]$ &$86.65\ [83.16-90.14]$ \\
causal SNP & $99.03\ [98.39 -99.67 ] $ &$100.0\ [100.0-100.0]$  & $92.01\ [89.13-94.89]$ &$99.08\ [98.47-99.69]$ \\
\hline
\end{tabular}
\end{center}
\caption{\label{tab-auc-treatment} Association analysis performed on Chromosome 6 with the four approaches, when the confounding factor $\tt{treatment}$ is considered known. Restricted regions are centered on causal SNP}
\end{table}

\newpage

\newpage

\end{document}